# Organ-specific Branching Morphogenesis

Christine Lang[1,2,†], Lisa Conrad[1,2,†], Dagmar Iber[1,2,*]

[1] Department of Biosystems, Science and Engineering (D-BSSE), ETH Zurich, Mattenstraße 26, 4058 Basel, Switzerland

[2] Swiss Institute of Bioinformatics (SIB), Mattenstraße 26, 4058 Basel, Switzerland

[†] These authors have contributed equally to this work and share first authorship.

**\* Correspondence:**
Dagmar Iber
dagmar.iber@bsse.ethz.ch

## Abstract

A common developmental process, called branching morphogenesis, generates the epithelial trees in a variety of organs, including the lungs, kidneys, and glands. How branching morphogenesis can create epithelial architectures of very different shapes and functions remains elusive. In this review, we compare branching morphogenesis and its regulation in lungs and kidneys and discuss the role of signaling pathways, the mesenchyme, the extracellular matrix, and the cytoskeleton as potential organ-specific determinants of branch position, orientation, and shape. Identifying the determinants of branch and organ shape and their adaptation in different organs may reveal how a highly conserved developmental process can be adapted to different structural and functional frameworks and should provide important insights into epithelial morphogenesis and developmental disorders.

## 1 Introduction

Highly branched tubular structures are a common architectural motif of many organs, such as lungs, kidneys, and glands, that require a high surface to volume ratio for their function (Lu and Werb 2008; Ochoa-Espinosa and Affolter 2012; Iber and Menshykau 2013; Spurlin and Nelson 2016; Goodwin and Nelson 2020). The central structure of branched organs is composed of tightly associated epithelial cells surrounded by rather loosely connected mesenchymal cells. The developmental process during which epithelial buds branch into the surrounding mesenchyme and thereby construct a complex epithelial tree is called branching morphogenesis. The biochemical and mechanical interactions between epithelium and mesenchyme that regulate branching morphogenesis have been studied in great detail, and the key regulatory factors that control branching morphogenesis are largely well known (Jones et al. 2021; Warburton et al. 2005; Shah 2004). How the commonly used principle of branching morphogenesis can create organs of different shapes and functions, nonetheless, remains elusive.

In the following, we will focus on the regulation of branching processes in the lung and kidney because both organs share many similarities with regard to branching morphogenesis, but still have a very different organ shape and function. For both organs, branching is suggested to be highly stereotyped, which does not apply to glandular organs like pancreas, prostate, thyroid, salivary and mammary gland (Villasenor et al. 2010; Lu and Werb 2008; Wang et al. 2017). In contrast to lung and kidney, branching patterns in the mammary gland or prostate are substantially affected by hormones (Sternlicht et al. 2006; Prins and Putz 2008). Epithelial branching in lungs and kidneys is accompanied by overall organ growth, while the stroma in the mammary gland is already established and does not substantially change in size when the epithelium starts to invade it (Sternlicht et al. 2006).

Branching morphogenesis has been studied in model organisms other than the mouse, rat and the chick (Affolter and Caussinus 2008; Horowitz and Simons 2008; Shifley et al. 2012; Rankin et al. 2015; Bracken et al. 2008). While orthologs of many genes involved in branching have similar roles in other species, these are not within the scope of this review.

## 2 Branching Morphogenesis in Lung and Kidney

Branching morphogenesis of the mouse lung primarily occurs during the pseudoglandular stage of lung development between embryonic day (E) 10.5 and E16.5 (Herriges and Morrisey 2014; Schittny 2017). At around E10.5, after initial bronchial formation, the lung lobes have started to form, with the right lung consisting of four lobes and the left lung remaining as a single lobe, and then continue to repeatedly elongate and branch into the surrounding mesenchyme (Cardoso and Lü 2006; Warburton et al. 2010). After 13-17 branch generations, the lung tree consists of more than 5000 branches (Metzger et al. 2008). Kidney branching morphogenesis in the mouse starts around E11.5 after the ureteric bud (UB) has invaded the metanephric mesenchyme (MM) and established the characteristic T-shape (McMahon 2016). Subsequent repeated rounds of around 9-12 generations of epithelial branching and elongation transform the UB into a complex tree with around 2600 branch and tip segments at E16.5 (Short et al. 2014; Cebrián et al. 2004).

Most branched organs require the strict modulation of at least one receptor tyrosine kinase (RTK) signaling pathway whose modulation is strictly controlled (Lu and Werb 2008). The associated ligands are usually expressed in the mesenchyme, while the respective receptors are produced in the epithelium. Fibroblast growth factor 10 (FGF10) and glial cell line-derived neurotrophic factor (GDNF) signaling are essential for branching in the lung and kidney, respectively (Min et al. 1998; Sekine et al. 1999; Moore et al. 1996; Pichel et al. 1996; Sánchez et al. 1996). FGF10 signals through its receptor fibroblast growth factor receptor 2b (FGFR2b), and GDNF via its receptor rearranged during infection (RET) and co-receptor GDNF family receptor α (GFRα) (Ohuchi et al. 2000; Jing et al. 1996; Treanor et al. 1996). *Fgf10^{-/-}* and *Fgfr2b^{-/-}* mice do not develop lungs and die at birth (Ohuchi et al. 2000; Moerlooze 2000), while the absence of GDNF or its receptors results in kidney agenesis (Schuchardt et al. 1994; 1996; Moore et al. 1996; Pichel et al. 1996; Sánchez et al. 1996; Cacalano et al. 1998). FGF10 and GDNF both induce extracellular-signal-regulated kinase (ERK) and phosphoinositide 3-kinase (PI3K) signaling and converge on the transcription factors *ETS translocation variant 4/5* (*Etv4/5*) to promote branching morphogenesis (Liu et al. 2003; Herriges et al. 2015; Lu et al. 2009; Zhu and Nelson 2012; Fisher et al. 2001; Tang et al. 2002). Another important shared signaling pathway is wingless-related integration site (Wnt)/β-catenin (De Langhe et al. 2005; Hashimoto et al. 2012; Ostrin et al. 2018; Sarin et al. 2014). FGF10 signaling plays an important role during kidney branching as well. Thus, kidneys from *Fgf10^{-/-}* and *Fgfr2^{-/-}* mice are characterized by reduced size and dysplasia (Ohuchi et al. 2000; Michos et al. 2010; Zhao et al. 2004). Moreover, treatment of embryonic kidney cultures with soluble FGFR2b inhibits branching (Qiao et al. 2001). In the absence of *Sprouty1*, an ERK signaling inhibitor, FGF10 rescues the *Gdnf^{-/-}* phenotype and is therefore considered to have an at least partly redundant function to GDNF (Michos et al. 2010). GDNF and its receptors are expressed in the lung as well, but only from E13.5 onwards, and neither exogenous GDNF nor a GDNF-blocking antibody affects lung development (Towers et al. 1998).

Both organs show specific expression patterns in epithelium and mesenchyme for FGF10 and GDNF and the respective receptors. During lung branching morphogenesis, *Fgf10* is expressed in spatially restricted domains in the submesothelial mesenchyme around distal epithelial buds, while the expression of its receptor *Fgfr2b* is localized to distal lung epithelia (Fig. 1A, Bellusci et al. 1997). At E13.5, *Gdnf* is present throughout the lung mesenchyme, while *Ret* is detected throughout the epithelium and its co-receptor *Gfra* is expressed exclusively in the proximal epithelium and mesenchyme (Towers et al. 1998). In the kidney, *Fgf10* is expressed throughout the MM at E11.0 and becomes enriched in the cap mesenchyme (CM) by E12.5, while *Fgfr2* expression is present in the UB epithelium (Fig. 1B, Michos et al. 2010; Trueb et al. 2013; Brunskill et al. 2014; Zhao et al. 2004; Sanna-Cherchi et al. 2013). *Gdnf* is expressed in the CM, which surrounds the branching UB tips, and in the kidney stroma (Fig. 1B, Hellmich et al. 1996; Magella et al. 2018). *Ret* is expressed exclusively in the UB and becomes restricted to bud tips through a cell sorting mechanism as branching morphogenesis progresses, while its co-receptor *Gfra1* is expressed in both the UB and the MM (Fig. 1B, Pachnis 1993; Riccio et al. 2016; Rutledge et al. 2017). The spatially restricted expression domains of FGF10 in the lung and RET in the kidney suggest that distinct expression patterns play an important role in regulating organ-specific branching. However, uniform *Fgf10* expression does not abrogate lung branching morphogenesis but regulates epithelial lineage commitment (Volckaert et al. 2013). Similarly, expression of *Ret* throughout the UB instead of only at the tips does not inhibit branching but causes retardation of kidney development (Srinivas et al. 1999).

FGF10 and GDNF signaling are regulated by several stimulatory and inhibitory factors, including SHH, Sprouty, WNT, and bone morphogenetic protein (BMP)/transforming growth factor-β (TGF-β) signaling (Yuan et al. 2018; Costantini and Kopan 2010). FGF10 signaling in the lung triggers the expression of *Shh* in the epithelium which in turn represses *Fgf10* expression in the mesenchyme (Fig. 1C, Bellusci et al. 1997; Lebeche et al. 1999). This negative feedback loop between FGF10 and SHH regulates the spatial-temporal dynamics of *Fgf10* expression in the lung mesenchyme. GDNF signaling in the kidney upregulates *Wnt11* expression in the epithelium, enhancing *Gdnf* expression in the mesenchyme (Fig. 1D, Majumdar et al. 2003). This positive feedback between GDNF and WNT11 plays an important role in the regulation of tip packaging in the kidney (Menshykau et al. 2019).

A further distinction can be made between the distal branch tips and the proximal stalks of the branching epithelium. Tips and stalks are characterized by distinct sets of expressed genes, some of which are specific to the lung or UB tip, while others are expressed in both organs (lung tip genes summarized in Nikolić et al. 2017; Rutledge et al. 2017).

## 3 Hallmarks of Branching

During organogenesis, lungs and kidneys generate highly arborized epithelial networks from simple epithelial organ bud primordia (M. Herriges and Morrisey 2014; Schittny 2017; McMahon 2016). The organs differ in the extent and rate of branching, i.e., the number of created branches after a certain amount of time, as well as in the distance, orientation, and branch shape (Fig. 1E). How these characteristics are regulated during branching morphogenesis in different organs is an important field of study.

Branching processes are typically studied using microscopy imaging data. Organ explant cultures have proven to be a valuable method to monitor branching processes in 2D over a desired period of time (del Moral and Warburton 2010; Michos 2012). This method, however, is not able to capture branching aspects in 3D and is reported to recapitulate *in vivo* branching only to a certain extent due to emerging artifacts (Watanabe and Costantini 2004; Short and Smyth 2016). 3D imaging provides comprehensive reconstructions of branched trees (Short et al. 2010). As live imaging of organs in 3D over long periods of time is still limited, only snapshots of certain developmental stages are available so far. The quantitative analysis of either 2D or 3D imaging data on branching morphogenesis is challenging due to the high complexity of branched structures. Image processing and analysis methods have now advanced over the last years to enable comprehensive studies on branch position, branch orientation, and branch shape (Short and Smyth 2016). All these characteristics are essential to describe and compare epithelial tree architectures (Fig. 1E). In this section, we highlight developmental mechanisms that are known to contribute to the regulation of these branching hallmarks.

## 3.1 Branch Position

*The sequence and stereotypy of branch emergence*
To arrive at an elaborately branched epithelial tree from a simple organ anlage, new branches need to emerge. Generally, new branches can be formed by the splitting of an existing tip in two or more daughter branches or by the formation of a new bud on the side of an existing tube. The sequence in which new branch points emerge is a significant determinant of the final branched tree and differs markedly between the lung and the kidney (Metzger et al. 2008; Short et al. 2013). The lung epithelium forms new branches by lateral branching and through tip bifurcations in planar or orthogonal orientation (Fig. 2A, Metzger et al. 2008; Short et al. 2013). In the kidney, new branches form through tip bifurcations and sometimes trifurcations, which, however, resolve into bifurcations during continued development (Fig. 2A, Short et al. 2014). Although the direction in which branches can grow is constrained by the flattened morphology of the organ, lungs and kidneys cultured in *ex vivo* organ culture systems form new branches using the same branching modes as their *in vivo* counterparts. However, in kidney explants, lateral branching events at 'trunk' segments are sometimes observed (Lin et al. 2003; Watanabe and Costantini 2004).

The branching program of the murine embryonic lung has been characterized in great detail by Metzger and colleagues, suggesting that lung branching is highly stereotyped (Metzger et al. 2008). The 3D reconstructions of E14.5 lungs from Short et al. correspond well with the previously proposed branching trees, although substantially more variations in branch positions and directions are observed, suggesting that stereotypy in lung branching is generally robust, but also allows for deviations on a finer scale (Short et al. 2013). The relaxation of the branching stereotypy beyond the first rounds of branching has been related to dynamic changes in the growth of the mesenchyme with branches adapting to the available space, highlighting the importance of coupled epithelial-mesenchymal growth (Blanc et al. 2012). A similar picture emerges for the kidney since internal branches appear to emerge in a highly stereotypical manner, while branches at the periphery show a higher degree of variations. This finding is based on subdividing the ureteric tree into several anterior and posterior clades, according to the first branches formed at E12.5, which show comparable branching patterns and complexity between individual organs and potentially manifest a comparable structural stereotypy as the different lobes in the lung (Short et al. 2014; Lefevre et al. 2017).

*The mesenchyme as a modulator of branching patterns*
Branching morphogenesis depends on inductive signals and complex, reciprocal interactions between the branching epithelium and the surrounding mesenchyme. The branching pattern of the lung epithelium is coupled to the shape of the surrounding mesenchyme. Dynamic changes in the direction of mesenchyme growth guide the specification of branch points and subsequent branch outgrowth. For example, domain branches are formed when the mesenchyme growth changes in a perpendicular direction during lung development (Blanc et al. 2012).

Tissue recombination experiments in the salivary gland and the lung have demonstrated the ability of the mesenchyme to support continued branching of the UB (Grobstein 1955; Kispert et al. 1996; Sainio et al. 1997). Notably, only bronchial mesenchyme supports continued branching of the lung epithelium (Spooner and Wessells 1970; Alescio and Dani 1971). While the lung epithelium is viable and able to form single buds when recombined with mesenchyme from the submandibular, salivary and mammary gland, as well as with mesenchyme from other sources (Spooner and Wessells 1970; Duernberger and Kratochwil 1980), further branching is only observed when the mesenchyme is renewed (Lawson 1983).

The identification of mesenchymal-derived morphogens required for branching morphogenesis of the lung and the kidney epithelium has made it possible to culture branching isolated epithelia embedded in extracellular matrix (ECM) gels. While requiring the appropriate growth factors, this has demonstrated an intrinsic ability of the epithelium to form new branches (Nogawa and Ito 1995; Bellusci et al. 1997; Qiao et al. 1999; Varner et al. 2015; Conrad et al. 2021). However, the branching pattern in these isolated cultures appears very different from the *in vivo* tree as the formation of new branches eventually stops, highlighting the importance of the mesenchyme for the modulation of the branching pattern and the maintenance of cellular niches.

A detailed morphometric characterization of the branching patterns of recombined lung and kidney tissue has shown that the lung mesenchyme 'reprograms' the UB to adapt a branching pattern that is typical for the lung. This demonstrates the importance of the mesenchyme in specifying the branching mode and in establishing organ-specific branching patterns (Fig. 2B; Lin et al. 2003). Similar reprogramming of organ-specific branching is observed when the mammary epithelium is recombined with salivary gland mesenchyme, suggesting a broader context in which the mesenchyme can specify epithelial branching patterns (Kratochwil 1969; Sakakura et al. 1976). In tissue recombination experiments, the mesenchyme typically gets cut up into pieces and placed around the epithelium, scrambling any pre-patterning. The use of organ-specific branching modes of lungs and kidneys is maintained in homotypic recombination experiments (Fig. 2B; Lin et al. 2003), suggesting that new branch points are formed spontaneously from regulatory interactions.

*A ligand-receptor-based Turing mechanism regulates branch point specification and distance*

The question of how the patterning and branching of epithelial tubes are regulated has inspired many different models for branching morphogenesis (reviewed in Iber and Menshykau 2013; Lang et al. 2018). What kind of mechanism could result in highly stereotyped branching, independent of a pre-patterning, and at the same time allow for variations in the branching pattern and adaptations to available space in the mesenchyme? Alan Turing's work on pattern formation in biological systems, which showed that self-organized symmetry breaks could result from diffusion-driven instabilities of morphogens, laid the foundation for computational studies on pattern formation (Turing 1952; Gierer and Meinhardt 1972). Our group has shown that the ligand-receptor interactions between FGF10 and FGFR2b, as well as between SHH and its receptor Patched-1 (PTCH) in the lung (Kurics et al. 2014; Menshykau et al. 2012; 2014) and between GDNF and RET in the kidney (Menshykau et al. 2019; Menshykau and Iber 2013) can give rise to Turing patterns (Fig. 1C,D). First modeled on an idealized tube domain, the interactions between FGF10, SHH and PTCH, as well as GDNF/RET signaling in a positive feedback loop with WNT11, both give rise to Turing patterns that, depending on the parameter values of the model, represent distinct branch modes in the lung and the kidney, namely tip bifurcation, trifurcation and lateral branching (Menshykau et al. 2012; Menshykau and Iber 2013). The frequency of the branch modes corresponds well to the *in vivo* situation, with the lung model favoring lateral branching and bifurcations and the kidney model predicting robust bifurcation and trifurcation patterns, but lateral branching to be rare (Menshykau et al. 2012; Menshykau and Iber 2013). The FGF10/SHH/PTCH-based model further predicts that the domain's growth speed influences branch mode selection, with faster growth leading to lateral branching events (Menshykau et al. 2012).

Given the dynamic nature of branching morphogenesis, a mechanistic model would additionally need to support branch outgrowth at the predicted branch points. Using time-lapse movies of lung and kidney 2D explant cultures and developmental sequences of 3D lung buds, our group has shown that only the ligand-receptor-based Turing model correctly reproduces the areas of branch outgrowth (Menshykau et al. 2014; 2019). Tissue-restricted expression of ligands and their receptors, as observed *in vivo* (Fig. 1A,B), makes the Turing model robust to noisy initial conditions, resulting in a geometry effect that can explain the stereotyped branching observed in the lung and the kidney (Menshykau et al. 2014; 2019; Metzger et al. 2008; Short et al. 2013). The ligand-receptor-based Turing mechanism predicts the highest signaling concentration at new branch points, which corresponds well to the *in vivo* pattern of pERK, a common readout of FGF10 and GDNF signaling (Chang et al. 2013; Ihermann-Hella et al. 2014; Conrad et al. 2021). Indeed, if components of the MAPK pathway are knocked out in the epithelium, no new branches can form beyond the main bronchi of the lung and the T-shape of the UB (Boucherat et al. 2015; Ihermann-Hella et al. 2014). Including the negative feedback from SHH/PTCH signaling in the FGF10/FGFR2b signaling-based model greatly increases the parameter space in which Turing patterns are observed, the so-called Turing space (Menshykau et al. 2014; Kurics et al. 2014). Likewise, the confinement of receptors to single cells and their tendency to form larger clusters greatly increases the Turing space (Kurics et al. 2014). A wider range of possible parameters allows for more variation in branching patterns and would make the evolution of a Turing mechanism more likely. Expanding the GDNF/RET-based Turing model with a positive feedback on *Gdnf* expression via WNT11 signaling improves the model fit to the experimental growth fields at later stages of kidney development when branch tips start to grow towards each other and compete for ligand (Menshykau et al. 2019). As predicted by the Turing mechanism, branches remain further apart in *Wnt11* null kidneys (Menshykau et al. 2019). Taken together, the ligand-receptor interactions likely represent the core patterning module that drives stereotypic branching morphogenesis. Coupled with other Turing modules and further feedbacks, a wide range of patterns can be robustly achieved to enable the different stereotypic branching patterns in the different organs.

The distance at which a new branch point is specified relative to other, already formed tips has an additional impact on the shape of the branched tree. A key prediction of the ligand-receptor-based Turing mechanism is that the spacing of branch tips depends on the rate of ligand expression and is only affected if this rate falls below a threshold (Celliere et al. 2012). Wider spacing of the first three lateral branches of the left lung lobe at E12.0 is observed in *Fgf10* hypomorphic lungs, but other allelic combinations with a milder reduction in *Fgf10* expression did not differ from wild type lungs (Ramasamy et al. 2007). A 55% reduction of *Fgf10* expression in hypomorphic lungs compared to the wild type is in line with the simulated threshold of a 50% decrease in expression rate (Fig. 2C) (Celliere et al. 2012; Ramasamy et al. 2007).

There is ample experimental evidence that biomechanical forces affect branching morphogenesis. Thus, alteration of the transmural pressure using microfluidic chambers or tracheal occlusion influences branching morphogenesis and maturation of embryonic lungs, with increased pressure leading to an increase in the formation of branches (Nelson et al. 2017; Unbekandt et al. 2008). Tracheal occlusion did not affect the final organ size, whereas the distance between branches was reduced at the culture endpoint (Unbekandt et al. 2008). Notably, the expression of *Fgf10* is increased by roughly 50%, linking the impact of the intra-luminal pressure to signaling and the Turing mechanism (Celliere et al. 2012). The changes in branch distance observed in these studies were inferred from static imaging data and do not provide information for the distance between branch points at the time of branch point specification. Live-imaging at high temporal resolution would circumvent this problem.

*Dynamic cell behaviors during bud formation*

While the genetic regulation of branch patterning is increasingly well understood, it is less clear how signaling translates to dynamic cell behaviors and which morphological processes are necessary for the formation and outgrowth of a new bud. While differential proliferation has been a longstanding candidate, there are conflicting results on its role as a driver of bud outgrowth and branching (Ettensohn 1985). Increased proliferation in branch tips has been observed for the kidney (Michael and Davies 2004; Riccio et al. 2016) and in distal versus proximal regions of the lung epithelium (Okubo et al. 2005). Proliferation was also shown to be restricted to already formed branch tips of mesenchyme-free lung epithelium cultures; however, proliferation was uniform before branching, arguing against differential proliferation driving bud formation *in vitro* (Nogawa et al. 1998).

In the lung, differential proliferation between tip and stalk was only observed for lateral branching but not during tip bifurcation (Schnatwinkel and Niswander 2013). However, blocking proliferation in the chicken lung, which forms new branches in a comparable fashion to mouse lateral branching, showed that proliferation is not required for the initiation of new buds (Kim et al. 2013). Furthermore, computational modeling demonstrated that proliferation alone does not support lung domain branching (Fumoto et al. 2017; Goodwin et al. 2019; Kim et al. 2013). Growth-induced mechanical instability could also result in the formation of new buds through epithelial buckling (Varner et al. 2015). In this model, differential proliferation in the epithelium and the surrounding mesenchyme (or matrigel) could support bud formation, yet spatial proliferation differences within a layer are not required.

Proliferation is required for tip enlargement into the characteristic ampulla in the kidney (Michael and Davies 2004; Ihermann-Hella et al. 2014). Nonetheless, no difference in proliferation between the bifurcating tip's lateral sides and the cleft has been observed in either the lung or the kidney, making it difficult to explain how increased tip proliferation alone could lead to bifurcations (Michael and Davies 2004; Schnatwinkel and Niswander 2013). Bifurcating tips in the lung showed differential cell division orientation at the newly forming tip versus the cleft region, with more divisions contributing to elongation in the cleft, potentially 'pushing' the daughter tips apart (Schnatwinkel and Niswander 2013). Luminal mitosis has been observed in the UB, during which one daughter cell loses its basal contact and reinserts into the UB a few cell diameters away (Packard et al. 2013), but it has not been demonstrated whether this process is needed for tip bifurcations and reinsertion seemingly happens in random locations.

Live-imaging of mosaic UBs has revealed a tip progenitor population that is dependent on *Ret* and the transcription factors *Etv4/5* (Kuure et al. 2010; Shakya et al. 2005). Tip cells compete for the tip domain based on the level of signaling and show high cell motility (Chi et al. 2009), but whether cells exhibit coordinated movement or directional migration that contributes to tip bifurcation remains to be shown. Genes downstream of *Etv4/5* are involved in cellular migration, adhesion, and ECM remodeling, which could, in principle, drive cell sorting mechanisms (Lu et al. 2009). GDNF has been shown to act as a chemotactic factor for the renal Madin-Darby canine kidney (MDCK) cell line. However, it is not required as a paracrine factor for UB patterning during kidney development through a localized source of GDNF (Shakya et al. 2005; Tang et al. 1998). Similar to *Ret* dependent cell sorting, $Fgfr2^{UB-/-}$ cells show less frequent occupation of UB tips as compared to wildtype cells, which might provide some redundancy of GDNF and FGF signaling for cell rearrangements in the UB (Leclerc and Costantini 2016). The matrix metalloproteinases (MMPs) MMP2 and MMP14 are localized at UB tips (Meyer et al. 2004; Pohl et al. 2000; Kanwar et al. 1999), and reduced mechanical resistance and signaling factors released by degradation of the ECM could facilitate tip outgrowth into the mesenchyme. Altering MMP activity in the lung can lead to decreased, as well as increased branching, presumably depending on the extent to which the ECM is altered. (Gill et al. 2003; Rutledge et al. 2019). For example, a low concentration of an MMP inhibitor enhances focal ECM deposits, which leads to increased branching, whereas high concentrations reduce branching (Gill et al. 2006).

Ectopic basolateral localization of E-cadherin in *Mek1/2* knockout UB and increased *E-cadherin* expression as a consequence of reduced FGF10 signaling in the lung suggests that increased cellular adhesion has an adverse effect on bud formation (Ihermann-Hella et al. 2014; Jones et al. 2019). Additionally, local fibronectin accumulation reduces E-cadherin in the apical cell-cell adhesion belt and fibronectin has been reported to support branching, suggesting that localized reduction of cellular adhesion allows epithelial remodeling processes that support bud bifurcation (Sakai et al. 2003). Clefting is particularly well studied in the salivary gland, and temporal imaging of fibronectin deposition in the progressing cleft has been shown to mediate bifurcation of branch tips (Larsen et al. 2006). Localized differentiation of mesenchymal-derived smooth muscle (SM) at the branch tip prior to cleft formation has been suggested to drive clefting of lung branches (Kim et al. 2015), but genetic inhibition of SM differentiation does not abrogate lung branching (Young et al. 2020).

Additionally, dynamic cell shape changes are observed during bud formation, with cells in the tips adopting a wedge-shaped morphology (Kadzik et al. 2014; Kim et al. 2013; Meyer et al. 2004). This could be achieved actively through apical constriction or be a consequence of the bud's geometry. Although this question has mainly been investigated in explant cultures supplemented with inhibitors of actomyosin contractility, studies have yielded conflicting results on the role of apical constriction (Kim et al. 2013; Michael, Sweeney, and Davies 2005; Moore et al. 2005; Schnatwinkel and Niswander 2013; Meyer et al. 2006). Interestingly, there seems to be a requirement for apical constriction for tip bifurcation, but not for lateral branching in the mouse lung (Schnatwinkel and Niswander 2013), while apical constriction is required for the initiation of new buds in the chicken lung (Kim et al. 2013).

### 3.2 Branch Orientation

*Stereotyped orientation of branching modes*

The first comprehensive analysis of branch angles in lungs and kidneys is provided by the morphometric studies of E14.5 lung and E15.5 ureteric trees by Short et al. (Short et al. 2013; 2014). Branching events in the lung occur by either domain branching or bifurcations (Metzger et al. 2008). While lateral branches in the E14.5 lung tree are oriented parallel at an angle of around 80°, bifurcations show a divergence angle of around 115° (Fig. 2A, Short et al. 2013). In the kidney, bifurcations constitute the dominant branching mode. The corresponding divergence angles in the E15.5 ureteric tree are reported to be around 100° (Fig. 2A, Short et al. 2013; 2014). Bifurcations do not exclusively occur in the same plane but can also be rotated relative to the previous bifurcation. In both organs, such rotations of bifurcation events show a dihedral angle of around 60 - 65° (Fig. 2A). Therefore, branch angles for bifurcations and rotations are remarkably similar between embryonic lungs and kidneys (Short et al. 2013; 2014). In the kidney, rotational angles are constant across branch generations and developmental stages between E12.0 and E16.5 (Short et al. 2013; 2014). Overall, not only branch position but also branch orientation appears to be stereotypic in lungs and kidneys. Since the morphometric studies by Short et al. are endpoint analyses at certain developmental stages, they do not provide information on branch orientation at the time point of branching (Short et al. 2013; 2014).

*Remodeling processes and regulation of branch orientation*

Branch orientation does not seem to be rigidly fixed but subject to spatial and temporal remodeling processes. Between E12.0 and E16.5, local bifurcation angles (starting direction of daughter branches) are constant across branch generations and developmental stages in the kidney, while global bifurcation angles (direction relative to terminal branch point) show substantial spatial and temporal dynamics (Short et al. 2013; 2014). The divergence angle of ureteric tips is close to 180°, which corresponds to the characteristic T-shape, but reduces to around 120°, producing a Y-shape, when the tips mature into branches (Sims-Lucas et al. 2009). While the kidney generally shows uniform growth between E12.0 and E16.5, ureteric tree and organ volume are static between E13.25 and E13.75 despite increasing tip number and density (Short et al. 2014). At around this time, compressive remodeling of the internal branches leads to an increased curvature of these internal branches in kidneys (Fig. 2D, Short et al. 2013; 2014). Concurrently, terminal branches move closer towards each other (Short et al. 2013; 2014). Branch tip packaging within the organ increases with developmental time and is accompanied by a concurrent decrease of inter-tip distances (Short et al. 2010). In the kidney, the density of tips at the organ surface correlates with branching angles, as ureteric trees from *Tgf-β2[+/-]* mice show increased tip distances as well as increased bifurcation angles between branching events compared to wildtypes (Short et al. 2010; 2013). Treatment of embryonic kidney cultures with TGF-β1 significantly alters branching angles (Bush et al. 2004). Furthermore, it has been shown that branch angles are controlled by the presence of other nearby branch tips in embryonic kidney cultures (Davies et al. 2014). Interestingly, branches never collide with each other and rather avoid close contact (Blanc et al. 2012; Miura and Shiota 2002). Ureteric branch tips try to maximize their distance to each other, but at the same time, the presence of neighboring tips changes the branching angle towards smaller values and therefore narrower configurations (Davies et al. 2014; Sims-Lucas et al. 2009). In the kidney, this self-avoidance is potentially guided by BMP7, since inhibition of BMP7 leads to branch collisions and small divergence angles between branches and branches bend away from BMP7-soaked beads (Davies et al. 2014). BMP7 is suggested to negatively affect the positive feedback of WNT11 on GDNF (Goncalves and Zeller 2011). It has been shown experimentally and computationally that the kidney-specific positive feedback of WNT11 on GDNF prevents depletion of GDNF, thereby allowing the dense packing of ureteric branch tips (Menshykau et al. 2019).

While the first generations of lung branching are highly stereotyped, the subsequent branching events show substantially more variations, also in the orientation of branches (Metzger et al. 2008; Blanc et al. 2012; Short et al. 2013). These variations are mainly observed in regions where spatial restrictions are less stringent, like wide-opened mesenchymal areas, suggesting that new branches grow homogeneously into the mesenchyme by following the main direction of mesenchyme growth and adapt their orientation to local changes in the mesenchyme shape (Blanc et al. 2012). How lung branches sense the mesenchymal growth pattern is still elusive. FGF10 has been proposed to act as chemoattractants in the lung since epithelial buds grow towards localized sources of FGF10 (Bellusci et al. 1997; Park et al. 1998). *In vitro*, lung endoderm extends towards FGF10-soaked beads independent of its association with mesenchyme (Park et al. 1998; Weaver et al. 2000). Consequently, the spatiotemporally dynamic expression pattern of *Fgf10* in the lung mesenchyme has been suggested to guide directional branch outgrowth during lung branching morphogenesis (Bellusci et al. 1997; Park et al. 1998). In addition, blood vessels have been reported to direct branch orientation in the lung as vascular ablation leads to an incomplete rotation of dorsal-ventral branches (Lazarus et al. 2011). Since vascular ablation results in alterations of the spatial expression patterns of *Fgf10*, *Shh* and *Sprouty2*, blood vessels may direct branch outgrowth by modulating the core signaling networks.

### 3.3 Branch Shape and Elongation

*Spatial and temporal dynamics of branch shape*
The length and width of branches in lungs and kidneys show substantial spatial and temporal dynamics during branching morphogenesis. In E15.5 kidneys, branch lengths and diameters decrease with increasing branch generation (Short et al. 2013). Similarly, E14.5 lungs present decreasing branch diameters with successive branch generation (Short et al. 2013). This trend is also observed when embryonic lung and kidney explants are cultured on a filter system (Conrad et al. 2021). Therefore, reducing branch diameters with progressive branch generations is likely a common characteristic of organ architecture in lungs and kidneys. Between E12.0 and E16.5, most branches of the ureteric tree increase in length and diameter over time, but not necessarily in a uniform way (Short et al. 2014). When ureteric tree and organ volume are static between E13.25 and E13.75, branching continues and branch tip length and volumes but not diameter are reduced (Short et al. 2014). A multitude of studies have queried perturbations on branch shape in both organs, but it is still unknown how these complex remodeling processes are regulated in detail. Interestingly, the shape of branch stalks and tips seem to be modulated differently (Fig. 2E).

*Biased cell division accompanies biased branch elongation*
The branches in both organs show anisotropic growth due to a biased mitosis spindle orientation. In the kidney, planar cell polarity (PCP) signaling is important for directing the orientation of cell divisions and tubule elongation (Saburi et al. 2008). For the lung, diverging reports exist on the role of the PCP pathway. Mutations in the PCP genes *cadherin EGF LAG seven-pass G-type receptor 1* (*Celsr1*) and *van gogh-like2* (*Vangl2*) result in a reduced number of epithelial buds, but larger branch widths in embryonic lung cultures, while FGF10 signaling and cell proliferation seem to be unaffected in mutant lungs (Yates et al. 2010). In contrast to that, Tang et al. did not find any differences in branch shape or spindle orientation in lungs homozygous for *Vangl2* (Tang et al. 2011). The authors rather show that the control of mitotic spindle orientation is linked to ERK1/2 signaling in the lung and that the bias in elongation is lost in mutants with a constitutively active form of Kirsten rat sarcoma viral oncogene (KRas) (Tang et al. 2011). However, independent of the involved pathway, the question arises where biased elongation originates from.

*Impact of localized signaling on branch shape and elongation*
In lungs and kidneys, FGF10 and GDNF signaling is restricted to branch tips and necessary for branch formation (Michos et al. 2010; Min et al. 1998; Moore et al. 1996; Pichel et al. 1996; Sekine et al. 1999; Conrad et al. 2021; Rozen et al. 2009; Sánchez et al. 1996). However, localized signaling is not required for biased elongation since pharmacological inhibition of FGFR signaling as well as inactivation of *Fgf10* or *Fgfr2* does not abrogate branch elongation in embryonic lungs (Conrad et al. 2021; Abler et al. 2009). Also, in the kidney, branch initiation and elongation seem to be regulated separately since trunks have been observed to transiently elongate without the contribution of tip-derived cells (Shakya et al. 2005), and deletion of *Mek1/2* or pharmacological inhibition of MEK1 abrogates budding but not elongation (Ihermann-Hella et al. 2014; Fisher et al. 2001; Watanabe and Costantini 2004). Similarly, hepatocyte growth factor (HGF) and TGF-β promote elongation at the expense of branching in embryonic UB cultures (Davies et al. 1995; Ritvos et al. 1995; Bush et al. 2004). At the same time, treatment of embryonic kidney cultures with TGF-β1 results in thicker UB stalks (Bush et al. 2004). In double knockout mice for *Gdnf* and *Sprouty1*, FGF10 signaling is sufficient for UB branching. However, the branch elongation rate varies between branches of cultured explants, resulting in an irregular branching pattern (Michos et al. 2010). Interestingly, inhibition of FGFR signaling or the PI3K pathway in embryonic lung cultures as well as inactivation of *Fgfr2* in the kidney leads to reduced branch widths, whereas the elongation bias of the branches is not affected (Conrad et al. 2021; Wang et al. 2005; Zhao et al. 2004; Sims-Lucas et al. 2009).

*Impact of the mesenchyme & ECM on biased outgrowth*
In the lung, SM wrapping around the epithelium starts to appear at around E11.5 and has been suggested to shape branches (Goodwin et al. 2019). However, *Myocardin* inactivation, which inhibits SM differentiation, has been shown to have no effect on branching morphogenesis in E14.5 lungs (Young et al. 2020). In the kidney, SM is only present at the ureter but not at the branches (Bush et al. 2006). Therefore, branch morphology in the lung and kidney cannot be controlled by SM. As lung and kidney epithelial branches elongate even in the absence of mesenchyme, both a force or a chemical source from the mesenchyme can be ruled out as general drivers of biased branch elongation (Nogawa and Ito 1995; Qiao et al. 1999; Varner et al. 2015; Conrad et al. 2021). Notably, UB branches elongate and thin less in the absence of mesenchyme, suggesting that the mesenchyme affects but is not required for biased elongation (Conrad et al. 2021).

The ECM is generated both by epithelial and mesenchymal cells and thus exists also in mesenchyme-free cultures. The ECM is mainly composed of laminin, collagen and fibronectin and is extensively remodeled during branching morphogenesis (Harunaga et al. 2015; Kyprianou et al. 2020). MMPs degrade ECM components and are therefore important regulators of ECM thickness. Since the ECM is thinner at branch tips, it facilitates the invasion of epithelial buds into the mesenchyme, suggesting that ECM remodeling might drive biased branch elongation. Deletion of *Adamts18*, an MMP-encoding gene with branch-tip enriched expression, leads to distinct phenotypes in lungs and kidneys (Rutledge et al. 2019). While *Adamts18*[-/-] lungs show reduced branching and shorter primary airways, *Adamts18*[- [/-] kidneys develop two ureters rather than one, while branching seems to be unaffected. *Adamts18* knockout embryos do not show any difference in *Fgf10, Shh, Bmp4* and *Sprouty2* expression. In contrast to *Adamts18* null lungs, *Adamts18*[+/-] lungs exhibit increased branch formation, suggesting a complex relationship between MMP levels and branching. UBs secrete endostatin (ES), a cleavage product of collagen XVIII, by MMP-driven degradation and bind ES along the stalks, but not at the branch tips. Interestingly, the presence of recombinant murine ES inhibits outgrowth and branching of embryonic rat UB cultured in Matrigel, resulting in short and widened branches. The presence of an ES-neutralizing antibody enhances UB outgrowth and branching, resulting in increased branch lengths (Karihaloo et al. 2001). However, despite the impact of these local ECM modulations on branch growth and shape, pharmacological inhibition of MMPs does not affect the elongation bias in embryonic lungs (Conrad et al. 2021).

*Cell-based simulations suggest a pulling rather than a constricting force to drive branch outgrowth*
As we have seen before, the mesenchyme and ECM are unlikely to drive the elongation of epithelial branches by the action of compressive forces. Moreover, cell-based tissue simulations of epithelial lung growth demonstrate that external constricting forces that lead to the bias in outgrowth observed in embryonic lung epithelium, do not yield the observed bias in cell shape and division (Stopka et al. 2019). Therefore, another mechanism has to be at play to generate the bias in outgrowth. Further cell-based tissue simulations revealed that a pulling force in longitudinal direction reproduces the observed bias in outgrowth as well as in cell shape and division (Conrad et al. 2021). Actin-rich protrusions at branch tips could, in principle, generate such a pulling force that drives biased elongation but are not observed at branch tips in lungs (Conrad et al. 2021).

*Fluid flow as a driver of biased branch outgrowth*
Branches in embryonic lungs and kidneys show largely collapsed tubular morphologies with narrow luminal spaces (Conrad et al. 2021). Simulations of tube collapse show that mechanical deformations would result in nonuniform stress and curvature patterns in the tube cross-section which would be inconsistent with uniform biased outgrowth (Conrad et al. 2021). Accordingly, actin density, that has been reported to be affected by external mechanical stimuli, is distributed relatively uniformly in collapsed lung branches (Hirata et al. 2008; Hayakawa et al. 2011; Shao et al. 2015; Conrad et al. 2021).
Recently, fluid flow inside branches has been quantitatively analysed during early lung development (Conrad et al. 2021). Flow-induced shear stress could act as a tangential force on the apical side of epithelial cells, which in turn, could drive elongating outgrowth. Cells are not directly deformed by shear stress but sense the stress via primary cilia and respond by adapting their cell shape (Jain et al. 2010; Galbraith et al. 1998; Weinbaum et al. 2011). For the narrow luminal spaces of the collapsed tubes and the estimated flow velocity, the estimated shear stress levels are well within the range that has been reported for cells to sense via primary cilia (Nauli et al. 2013; Flitney et al. 2009; Resnick and Hopfer 2007). Therefore, shear stress acting in the longitudinal direction on the apical side of branches is able to explain biased elongation as well as cell division and may play an essential role in driving elongating outgrowth in lungs and kidneys (Conrad et al. 2021).

## 3.4   Branch Tip Shape

*Regulation of branch tip shape by core signaling networks*
Besides the shape and elongation of branch stalks, the bud tips are highly sensitive to pharmacological treatments and mutations. FGF10 and GDNF signaling are essential for growth and branching in lung and kidney, respectively, but are linked to branch shape regulation as well. FGF7, an alternative FGFR2b ligand, inhibits branching and results in dilated bud phenotypes in lung cultures (Cardoso et al. 1997; Simonet et al. 1995; Park et al. 1998; Tichelaar et al. 2000). While FGF10 triggers receptor recycling and cell migration, FGF7 induces receptor degradation and cell proliferation (Francavilla et al. 2013). FGF9 treatment leads to lung bud dilation and up-regulation of *Fgf10* expression in the mesenchyme (del Moral et al. 2006; White et al. 2006; Yin et al. 2011; Yin and Ornitz 2020). The effect of FGF9 on lung buds is observed even if *Mek* is deleted, suggesting that FGF9 mediates its effect via MEK-independent signaling pathways (Boucherat et al. 2015). It has been recently demonstrated that FGF9 signals via FGFR3, activates PI3K and thereby promotes distal epithelial fate specification and inhibits epithelial differentiation in the lung (Yin and Ornitz 2020). Hyperactive *Kras* and pharmacological inhibition of MEK, both downstream of FGFR2 signaling, induce dilated branch tips in the lung (Chang et al. 2013; Fisher et al. 2001; Boucherat et al. 2015; Yin and Ornitz 2020). In the lung, FGF10 is engaged in a negative feedback with SHH, meaning that FGF10 signaling induces *Shh* expression, which in turn represses *Fgf10* expression (Fig. 1C, Bellusci et al. 1997; Lebeche et al. 1999). Inhibition of SHH signaling by cyclopamine results in an increased number of small distal buds in lung cultures, while the stalks show an inflated phenotype (White et al. 2006).

In kidney cultures, adding GDNF either uniformly or loaded on beads leads to bud widening (Sainio et al. 1997; Pepicelli et al. 1997; Menshykau et al. 2019; Shakya et al. 2005). Dilated tips are also observed in UBs in which increased *Gdnf* expression is restricted to cells that naturally transcribe *Gdnf* (Li et al. 2019). This phenotype is reversed by pharmacological MEK inhibition and is suggested to be a result of restricted emigration of cells from tip to stalk regions and shortened cell-cycle time in tip cells (Li et al. 2019). Deletion of the negative regulator *Sprouty1* increases branching and *Gdnf* expression and results in swollen UB tips at the same time (Basson et al. 2005; 2006; Michos et al. 2010). Loss of phosphatase and tensin homolog (PTEN), which antagonizes the PI3K signaling pathway, leads to dilated UB tips, strongly resembling the bud morphology reported for GDNF excess or expression of a constitutively active form of RET (Kim and Dressler 2007; Sainio et al. 1997; Pepicelli et al. 1997; Menshykau et al. 2019; de Graaff 2001).

*Regulation of branch tip shape by ECM remodeling*

The ECM has been shown to influence branch morphology in lungs, kidneys and salivary glands. Treatment with an anti-laminin antibody or collagenase reduces lung branching and leads to dilatation of branches (Schuger 1991; Ganser et al. 1991; Miura and Shiota 2002). Collagenase treatment results in a dilated phenotype in the kidney and salivary gland, but not in the pancreas (Wessells and Cohen 1968). In the lung, dysregulation of *Sox9* expression induces cyst-like structures at branch tips and defects in laminin and collagen disposition (Rockich et al. 2013; Chang et al. 2013). In the lung, kidney, and salivary gland, fibronectin inhibition blocks cleft formation and branching, resulting in enlarged buds, while supplementation with exogenous fibronectin enhances branching morphogenesis (Sakai et al. 2003; De Langhe et al. 2005). Notably, accumulation of fibronectin in cleft regions is accompanied by an adjacent loss of E-cadherin localization, suggesting that fibronectin might regulate branching morphogenesis by converting cell-cell adhesion into cell-matrix adhesions (Sakai et al. 2003; Onodera et al. 2012). Moreover, fibronectin is suggested to be involved in WNT-regulated morphogenetic processes as inhibition of WNT signaling in the lung by Dickkopf-1 (DKK1) treatment results in decreased fibronectin deposition, impaired branching, dilated end buds, decreased SM actin expression and defects in vasculature formation (De Langhe et al. 2005). The expression of *Fgf10*, *Bmp4* and *Shh* is unaffected upon DKK1 treatment, indicating that the observed dilated branch shape is not due to the altered expression of these key regulatory molecules in lung branching morphogenesis. Notably, FGF9, which induces bud dilation in lung cultures, has been reported to increase *Dkk1* expression (del Moral et al. 2006).

During branching morphogenesis, the ECM is extensively remodeled by the degrading action of MMPs (Harunaga et al. 2015; Kyprianou et al. 2020). Null mutation for tissue inhibitor of metalloproteinases 3 (TIMP-3) leads to increased activation of MMPs, enhanced fibronectin degradation and dilated buds in lungs (Gill et al. 2003; 2006). Similarly, treatment of embryonic lungs with an MMP inhibitor at high concentrations impairs branching and induces the dilation of tips (Gill et al. 2003; 2006). Interestingly, low concentrations of the MMP inhibitor result in enhanced growth and branching (Gill et al. 2003; 2006). Epidermal growth factor (EGF), TGF-α and FGF7 induce dilation of branch tips and are suggested to induce MMP activity in the lung (Ganser et al. 1991; Miura and Shiota 2002). Moreover, efficient RTK signaling of FGF10 and GDNF requires sulphated glycosaminoglycans (GAG) that are expressed in the ECM. Inhibition of GAG synthesis by sodium chlorate disrupts branching morphogenesis in lungs and kidneys (Davies et al. 1995; Kispert et al. 1996; Michael et al. 2005; Shannon et al. 2003). In the lung, heparan sulphate proteoglycans are important for FGF10 binding to the distal epithelium and treatment with heparinase or oversulphated heparins leads to dilated lung buds (Izvolsky et al. 2003).

*Regulation of branch tip shape by cytoskeleton dynamics*

Epithelial cell shape and tension dynamics mediated by the actin-myosin cytoskeleton have been shown to play an important role during branching morphogenesis (Moore et al. 2005; Kadzik et al. 2014). In the kidney, chemical inhibition of myosin ATPase, disruption of actin microfilaments, or inhibition of rho-associated protein kinase (ROCK) inhibits branching and results in bloated and misshapen bud tips (Michael et al. 2005). While chemical inhibition of myosin ATPase or disruption of actin microfilament integrity block growth and branching in the lung, but do not seem to affect tip morphology, suppression of myosin light chain (MLC) phosphorylation by pharmacological ROCK or MLC kinase inhibition leads to a lung phenotype with dilated buds (Moore et al. 2005; Kadzik et al. 2014). In the lung, ROCK inhibition disrupts differential growth patterns, actin and ECM remodeling as well as the organization of vascular architecture and epithelial morphology (Moore et al. 2005). Wnt/Frizzled-2 (Fzd2) signaling is required for controlling cell shape changes by regulating ras homolog family member A (RhoA), which in turn activates ROCK. Deletion of the Wnt receptor *Fzd2* results in decreased RhoA activity, decreased phosphorylated MLC levels at apical cell surfaces as well as dilated distal buds in lungs (Kadzik et al. 2014). Notably, these *Fzd2* lung mutants show elevated and expanded *Fgf10* expression, which potentially contributes to the overall phenotype, while the expression patterns of *Fgfr2b*, *Shh* and *Bmp4* are not altered.

*Regulation of branch tip shape by cellular transport components*

Epithelia of developing lungs, kidneys, and salivary glands express several types of voltage-dependent calcium channels (VDCCs) that have been linked to branching and branch shape. Inhibition of VDCCs by nifedipine treatment leads to abrogated branching, dilated buds and reduced ERK activation, suggesting a link between VDCC activity and ERK phosphorylation (Kim et al. 2015; 2018). In the kidney, removal of a subset of claudins, essential components of tight junctions and important for paracellular transport, leads to a decrease in ureteric tip formation, enlarged bud tip lumens and a decreased complexity regarding tip morphology (El Andalousi et al. 2020).

*Regulation of branch tip shape by vasculature*

Interestingly, blood vessels have been shown to influence branch shape in the lung. Disruption of vascular assembly by overexpression of vascular endothelial growth factor-A (VEGF-A) decreases branching and dilates branch tips in the lung (Akeson et al. 2003). Similarly, vascular inhibition leads to dilated airway branches and reduced branching, while cell proliferation remains unaffected (Lazarus et al. 2011). Notably, vascular ablation alters the expression pattern of *Fgf10* and the expression levels of *Shh* and *Sprouty2*. The resulting perturbed branching phenotype resembles the phenotype caused by ectopic *Fgf10* expression.

*Regulation of branch tip shape by RNA molecules*

Finally, RNA molecules have been reported to affect branching morphogenesis (Yu 2014). Two microRNAs (miRNAs) that have originally been documented to regulate angiogenesis, the pro-angiogenic miR-221, and the anti-angiogenic miR-130, show similar opposing effects on branch tip morphology in the lung (Mujahid et al. 2013). While downregulation of miR-221 enhances branching and vascular network formation but decreases tip width, anti-miR-130 treatment results in reduced branching, a poorly developed vascular network, and dilated branch tips (Mujahid et al 2013). Accordingly, upregulation of these miRNAs triggers the corresponding inverse effect on lung branching, vascular network formation and tip morphology (Mujahid et al. 2013). Notably, one target of miR-221 is the Hox gene *Hoxb5* which, in turn, has been reported to affect the expression of *tenascin-C* in the ECM as well as the spatial restriction of *Fgf10* expression in the lung mesenchyme (Volpe et al. 2007).

Dicer is an endoribonuclease that is expressed in the lung at the onset of branching morphogenesis and processes miRNA and small interfering RNA into their mature forms, which in turn regulate gene expression. Epithelial inactivation of *Dicer* results in the arrest of new branch formation and the dilation of distal tips, while epithelial growth is unaffected (Harris et al. 2006). Moreover, *Dicer* null-mutants show increased *Fgf10*, *Sprouty2* and *Bmp4* expression. In the kidney, epithelial removal of *Dicer* leads to the formation of dilated cysts, disrupted branching morphogenesis, and the downregulation of *Wnt11* and *Ret*, while *Gdnf* expression is unaffected (Nagalakshmi et al. 2011).

## 4    Conclusion & Outlook

In this review, we compared branching morphogenesis in lungs and kidneys on the branch-level by focusing on developmental mechanisms that determine branch position, orientation, and shape. In both organs, epithelial-mesenchymal signaling interactions, but also the ECM, and the cytoskeleton appear to play crucial roles in regulating these hallmarks of branching. However, the presented findings are primarily based on the analysis of individual snapshots during development or qualitative descriptions of perturbed branching morphogenesis, meaning that data on branching characteristics at the time point of branch formation is mostly missing. As such, quantitative time-lapse analyses of branch position, orientation, and shape during lung and kidney development are needed to gain further insights into the regulation of organ-specific branching morphogenesis.

Tissue recombination experiments have demonstrated the mesenchyme's ability to establish organ-specific branching patterns (Lin et al. 2003). It is not unlikely that the mesenchyme modulates additional determinants of the branched tree, like the angles between parent and daughter branches and the final branch shapes, although these aspects have not been quantified in tissue recombination experiments. Comprehensive analyses of the impact of the mesenchyme on branching hallmarks are thus required to elucidate further the role of the mesenchyme in shaping epithelial trees in different organs.

Besides establishing a highly branched epithelial tree, other developmental processes are equally essential for shaping organs. Already during branching morphogenesis, the functional units start to form, i.e., the acini in the lung and the nephrons in the kidney (Schittny 2017; McMahon 2016), accompanied by highly complex remodeling processes. As a consequence, unlike the epithelial tree structure that is laid down during lung branching morphogenesis, the fully developed adult lung exhibits a fractal-like architecture (Weibel 1991). In the kidney, nephron formation comprises remodeling processes such as the elongation and tight packaging of unbranched tubules (Little et al. 2010). Combining live imaging, quantitative image analysis, and mathematical modeling will provide essential insights into the regulation of these remodeling processes and their contribution to organ-specific development.

## 5    Conflict of Interest

The authors declare that the research was conducted in the absence of any commercial or financial relationships that could be construed as a potential conflict of interest.

## 6    Author Contributions

CL and LC wrote, DI conceived and edited the manuscript. All authors contributed to the article and approved it for publication.

## 7    Acknowledgments

We thank members of the CoBi group for their feedback and discussions.

## 9    Figures

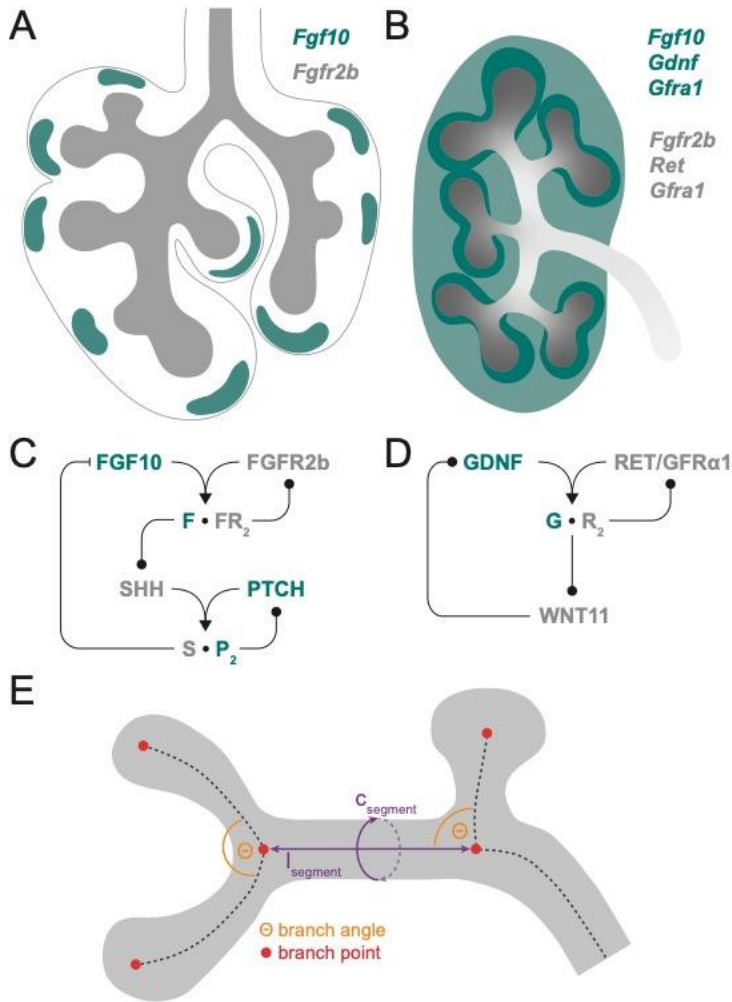

**Figure 1: Branching morphogenesis**

**(A)** Gene expression domains of *Fgf10* and its receptor *Fgfr2b* in the E11.5 mouse lung. *Fgf10* is expressed in a spotty pattern in the submesothelial mesenchyme (green), while *Fgfr2b* is expressed in the epithelium (grey).

**(B)** Gene expression domains of *Gdnf* and *Fgf10* and their receptors in the E12.5 murine kidney. *Gdnf* is expressed in both the cap mesenchyme (dark green) and the stroma (light green), while *Fgf10* is expressed in the cap mesenchyme. *Fgfr2b* is expressed in the ureteric bud (UB), while *Ret* expression is restricted to UB tips (dark grey). The *Ret* co-receptor *Gfra1* is expressed both in the metanephric mesenchyme and the ureteric bud.

**(C)** Core signaling network in lung branching morphogenesis. FGF10 signals via FGFR2b which enhances *Shh* expression. SHH signaling negatively regulates *Fgf10* expression.

**(D)** Core signaling network in kidney branching morphogenesis. GDNF signals via RET and GFRα1 which enhances *Wnt11* expression. WNT11 signaling positively regulates *Gdnf* expression.

**(E)** Hallmarks of branching: The morphology of branched epithelial trees is governed by the branching mode/sequence (branch points; red), the length and circumference of branches (purple) and by the branching angle (orange).

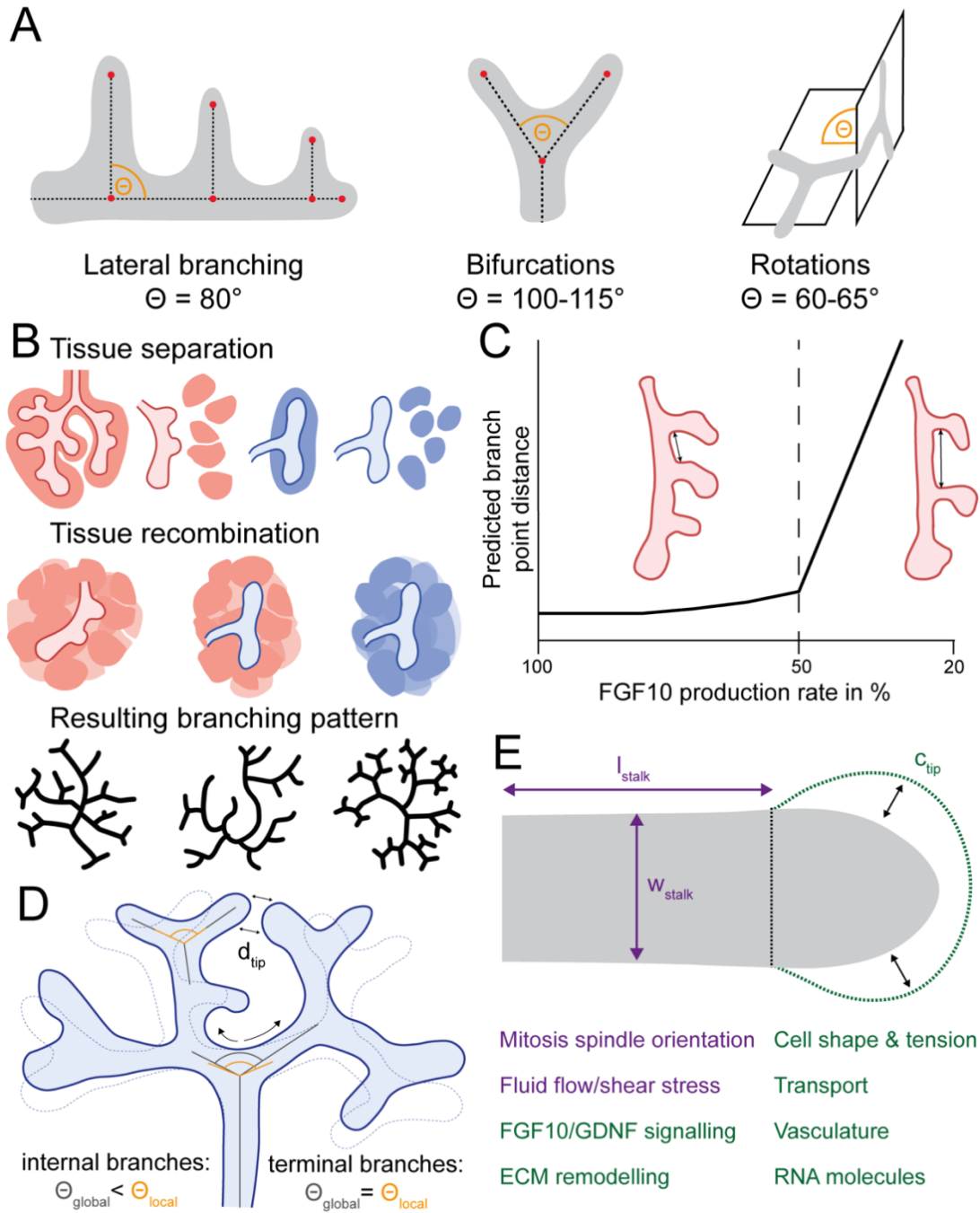

**Figure 2: Regulation principles of branching hallmarks in lung and kidney morphogenesis.**
(**A**) Branching modes used by lungs and kidneys. Lateral branches in the lung are oriented at an angle of around 80°. In both organs, bifurcations occur at a divergence angle of around 100 - 115°, while rotations of bifurcation events show a dihedral angle of around 60 - 65°.
(**B**) Mesenchyme as pattern modulator. Separation of mesenchymal and epithelial tissue and subsequent re-arrangement of the mesenchyme around the epithelium destroys any potential mesenchymal pre-patterning. In homotypic recombination experiments, the respective organ-specific branching pattern is maintained, while in heterotypic recombination experiments lung mesenchyme 'reprograms' the ureteric bud to adapt a lung-like branching pattern.

**(C)** Regulation of branch point distance in the lung. *Fgf10* hypomorphic lungs with a reduction of FGF10 expression by 55% exhibit a wider spacing of the first three lateral branches of the left lung lobe (right lung scheme), while other allelic combinations with a milder reduction in *Fgf10* expression do not differ from wild type lungs (left lung scheme). The lung schemes were reproduced from (Ramasamy et al. 2007). Accordingly, the ligand-receptor-based Turing mechanism predicts that the spacing of branch tips depends on the rate of ligand expression and is only affected if this rate falls below a certain threshold (dashed line). The graph was reproduced from (Celliere et al. 2012).

**(D)** Branch angle remodeling in the kidney. Local bifurcation angles (starting direction of daughter branches) are relatively constant, while global bifurcation angles (direction relative to terminal branch point) show spatial and temporal dynamics. Compressive remodeling of internal branches leads to an increased curvature of these internal branches. Concurrently, terminal branches move closer towards each other which reduces the inter-tip distance $d_{tip}$, thereby promoting tip packaging.

**(E)** Branch shape regulation. The shape of branch stalks and tips are regulated differently. The branches in lungs and kidneys show anisotropic growth, meaning that the increase in stalk length $l_{stalk}$ is larger than in stalk width $w_{stalk}$, due to a biased mitosis spindle orientation. Fluid flow and resulting shear stress is able to explain biased elongation in lungs and kidneys. Branch tip shape is regulated by several factors, such as signaling interactions, ECM remodeling or cell tension dynamics, which are highly interconnected. Perturbation of these shape determinants primarily leads to dilated buds characterized by an increased tip circumference $c_{tip}$ at the absent cleft formation.